\documentclass[a4paper]{article}

\usepackage{icrc2013}

\title{The ASTRI SST-2M Prototype: Camera and Electronics}

\shorttitle{ASTRI SST-2M Prototype: Camera and Electronics}

\authors{
Osvaldo Catalano$^{1}$,
Salvo Giarrusso$^{1}$,
Giovanni La Rosa$^{1}$,
Maria Concetta Maccarone$^{1}$,\\
Teresa Mineo$^{1}$,
Francesco Russo$^{1}$,
Giuseppe Sottile$^{1}$,
Domenico Impiombato$^{1}$,
Giovanni Bonanno$^{2}$,
Massimiliano Belluso$^{2}$,
Sergio Billotta$^{2}$,
Alessandro Grillo$^{2}$,
Davide Marano$^{2}$,\\
Vincenzo De Caprio$^{3}$,
Mauro Fiorini$^{4}$ and
Luca Stringhetti$^{4}$
for the ASTRI Collaboration$^{5}$, \\
Salvo Garozzo$^{2}$, Giuseppe Romeo$^{2}$\\
}

\afiliations{
$^1$ INAF - IASF Palermo, Via U. La Malfa 153, I-90146 Palermo, Italy \\
$^2$ INAF - Osservatorio Astrofisico di Catania, Via S. Sofia 78, I-95123 Catania, Italy \\
$^3$ INAF - Osservatorio Astronomico di Capodimonte, Salita Moiariello 16, I-80131 Napoli, Italy \\
$^4$ INAF - IASF Milano, Via E. Bassini 15, I-20133 Milano, Italy \\
$^5$ http://www.brera.inaf.it/astri\\
}

\email{Osvaldo.Catalano@iasf-palermo.inaf.it}

\abstract{ASTRI is a Flagship Project financed by the Italian Ministry of Education, University and Research, and led by INAF, the Italian National Institute of Astrophysics. The primary goal of the ASTRI project is the realization of an end-to-end prototype of a Small Size Telescope for the Cherenkov Telescope Array. The prototype, named ASTRI SST-2M, is based on a completely new double mirror optics design and will be equipped with a camera made of a matrix of SiPM detectors. Here we describe the ASTRI SST-2M camera concept: basic idea, detectors, electronics, current status and some results coming from experiments in lab.}


\begin{document}
\maketitle

\section{Introduction}

The Italian participation to the Cherenkov Telescope Array, CTA \cite{bib:CTA}, is mainly represented by the ASTRI program \cite{bib:ASTRI}, a ~"Flagship Project" financed by the Italian Ministry of Education, University and Research, and led by INAF, the Italian National Institute of Astrophysics. Primary goal of the ASTRI program is the realization of an end-to-end prototype of a Small Size Telescope \cite{bib:Pareschi} devoted to the highest gamma-ray energy region ($\sim1-100TeV$) investigated by CTA. The telescope, named ASTRI SST-2M, is \-cha\-rac\-te\-ri\-zed by innovative technological solutions for the first time adopted together in the design of Cherenkov telescopes: the optical system is arranged in a dual-mirror configuration \cite{bib:Canestrari,bib:Bonnoli} and the camera at the focal plane is composed of a matrix of multipixel Silicon Photo Multipliers, SiPM.

The ASTRI SST-2M prototype will be tested on field in Italy: the installation is foreseen in mid-2014 at the INAF "M.G. Fracastoro" observing station in Serra La Nave \cite{bib:Maccarone} near Catania. The data acquisition will start immediately after so to probe the technological solutions adopted and to verify the telescope expected performance \cite{bib:Strazzeri,bib:Bigongiari}.

\begin{figure}[h]
  \centering
  \includegraphics[width=0.44\textwidth]{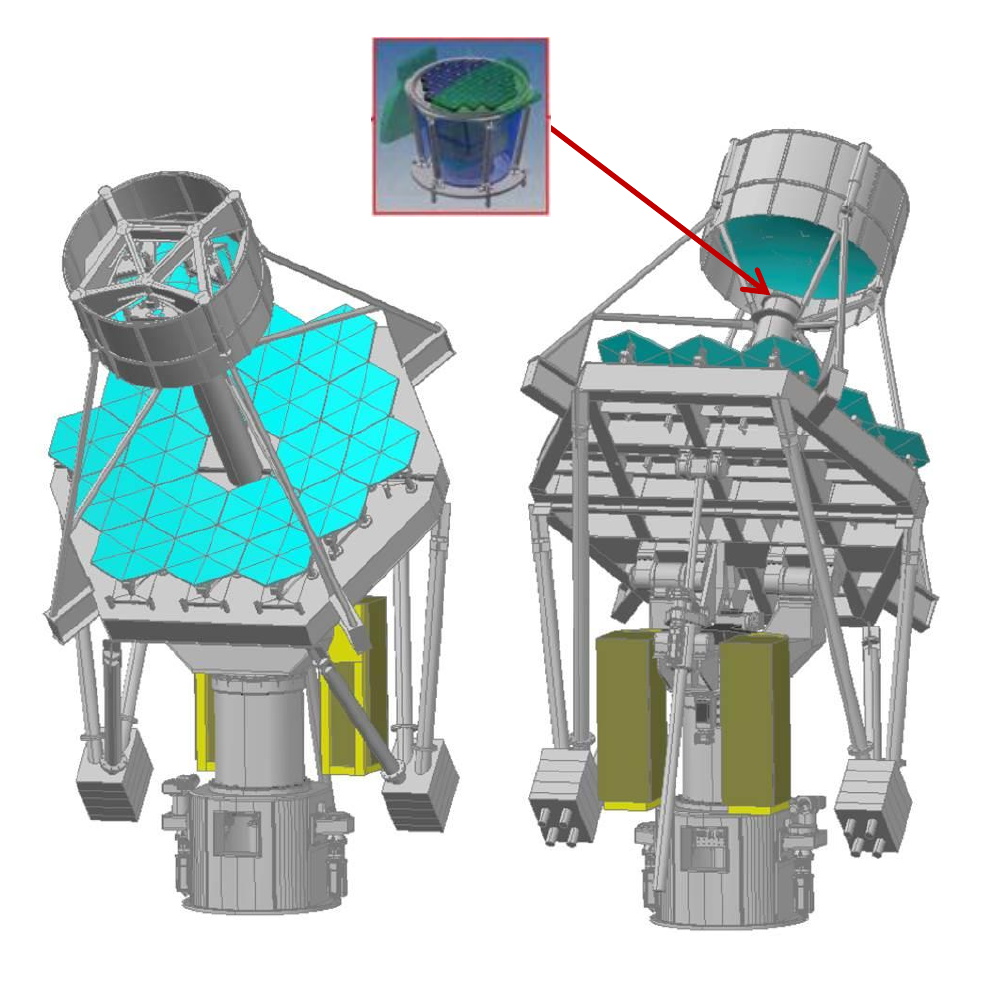}
  \caption{Layout of the ASTRI SST-2M telescope and of the camera with its protective lid.}
  \label{figure.1}
 \end{figure}

A sketch of the entire ASTRI SST-2M telescope is shown in Figure \ref{figure.1}. The telescope mount is of the alt-azimuthal type; the optical design, widely described elsewhere in these proceedings \cite{bib:Canestrari}, is based on a dual-mirror Schwarzschild-Couder configuration with f-number $f/0.5$, plate scale of $37.5mm/^\circ$, pixel size of approximately $0.17^\circ$, equivalent focal length of $2.15m$, and full field of view (FoV) of $9.6^\circ$.

This optical configuration allows to design a compact and lightweight camera to be placed at the curved focal surface of the telescope. The ASTRI SST-2M camera has a truncated-cone shape and its overall dimensions are about $560mm\times560mm\times560mm$, including mechanics and interfaces with the telescope structure, for a total weight of $\sim50kg$. The camera is composed by a matrix of SiPM which are suitable for the detection of the Cherenkov flashes being very fast and sensitive in the $300-700nm$ region.

In the following we will describe the main features of the ASTRI SST-2M camera design, its sensors and electronics, as well as some results coming from lab experiments.

\section{The ASTRI SST-2M Camera}

The SiPMs used for the ASTRI SST-2M camera becomes to a family of light sensors with very interesting characteristics. They are basically an array of Avalanche Photo Diodes working in Geiger-mode (GAPD) in which the reverse bias voltage is set beyond the Breakdown Voltage (over-voltage). In this way, a single photon absorbed in Silicon develops a saturated current avalanche with a gain of the order of $10^{6}$.

There are many advantages in using SiPMs compared to the traditional Photo-Multiplier Tubes: excellent single photon resolution, high Photon Detection Efficiency, low bias voltage (of the order of $30-90V$), no damage when exposed to ambient light, insensitivity to magnetic fields, small size. The drawbacks however are: high dark counts, after-pulses, optical crosstalk, gain strongly dependent on temperature. Nevertheless, all these negative features can be maintained under control: due to our pixel FoV, the rate of the dark counts is well below the rate of the Night Sky Background (22 MHz/pixel, corresponding to dark sky far from the Galactic Plane) so that the instrumental background does not degrade any further the telescope sensitivity; the effects of crosstalk and after-pulses are lower than $20\%$ (as shown in the following); the gain can be kept constant with temperature ($16^\circ$C) control and a proper setting of over-voltage.

 \begin{figure}[t]
  \centering
  \includegraphics[width=0.43\textwidth]{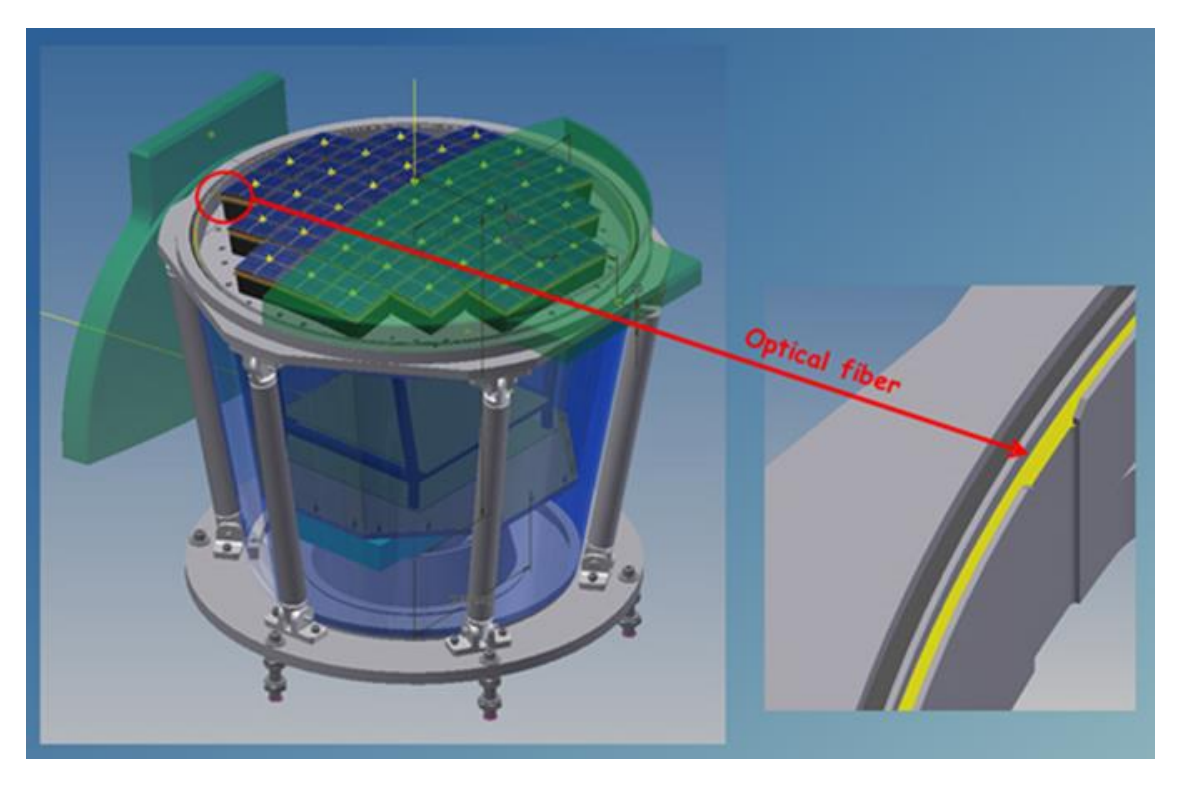}
  \caption{Schematic view of the ASTRI SST-2M camera box with its inner present fiber system for calibration purposes. On the top of the camera are visible the 37 PDMs.}
  \label{figure.2}
 \end{figure}

The small detection surface (about $500mm\times500mm$) of the ASTRI SST-2M camera requires a spatial segmentation of a few square millimeters to be compliant with the imaging resolving angular size ($0.17^\circ$). Among the \-a\-vai\-la\-ble SiPM sensors  we selected the Hamamatsu Silicon Photomultiplier S11828-3344M \cite{bib:Hamamatsu} working at a nominal $V_{bias }$ of $70V$; each sensor unit is formed by 4$\times$4 squared pixels, $3mm\times3mm$, made up of $3600$ elementary diodes of $50\mu$m pitch giving a filling factor of $62\%$.  In order to match the angular resolution of the optical system, in the design of the ASTRI SST-2M camera we used a modular approach: the physical pixels of each sensor unit are grouped in a $2\times2$ logical pixel ($6.2mm\times6.2mm$) having a sky-projected angular size of $0.17^\circ$. The aggregation of $4\times4$ sensor units ($8\times8$ logical pixels) forms the Photon Detection Module (PDM) and 37 PDMs (about 2000 logical pixels) form the camera at the focal plane;  this setup delivers the required full FoV of $9.6^\circ$. The advantage of this design is that PDMs are physically independent of each other, allowing maintenance of small portions of the camera. To fit the curvature of the focal surface, each PDM is appropriately tilted with respect to the optical axis of the telescope.

In order to protect the camera sensors from the external atmospheric environment, an optical-UV transparent, down to $300nm$, PMMA (Poly Methyl MethAcrylate) window is mounted on the PDM support structure. The PMMA window ($3mm$ thickness) is modeled with the same radius of curvature of the focal surface.
The ASTRI SST-2M camera is equipped with a light-tight lid (composed of two "petals" mounted onto the backbone structure of the camera) in order to prevent accidental sunlight exposure of the focal surface detectors, catastrophic in case of direct light reflected by the mirrors.
Eventually, between the PMMA window (along its inner circumference) and the upper surface of the backbone structure of the camera, will be allocated an optical fiber illuminated by a LED to perform relative gain calibration on the SiPM sensors. The camera is thermally controlled through a thermoelectric system based on Peltier cells. The thermal system is a Direct-to-Air thermoelectric assembly that uses impingement flow to transfer heat; it offers dependable, compact performance by cooling the camera via conduction.  The thermal system assures a proper working temperature to the camera electronics and in particular maintains the SiPM sensors at a constant temperature within $1^\circ$C. A sketch of the ASTRI SST-2M camera is shown in Figure \ref{figure.2}.

\section{The Electronics System}

The ASTRI SST-2M electronics assembly includes the SiPMs, the Front-End Electronics (FEE) and the Back-End Electronics (BEE). The focal surface is composed of 37 PDM tiles, each one hosting sixteen SiPMs with associated FEE.  The function of the FEE is to process and convert the analog SiPM signals into digital counts.  The FEE uses two Application Specific Integrated Circuits (ASIC) that interface the SiPM detectors, four Analogue-to-Digital Converters (ADC) and one Field Programmable Gate Array (FPGA) that manages the ASICs slow control, the generation of local trigger and the data acquisition.
The BEE is hosted on a separate common board.
It is based on a FPGA that, integrating a feature-rich dual-core processor, controls and manages the overall system, including data management, lid mechanisms and optical fiber calibration tool.
The BEE also provides the functions necessary to process and transmit the event data as obtained by the FEE.

 \begin{figure}[t]
  \centering
    \includegraphics[width=0.45\textwidth]{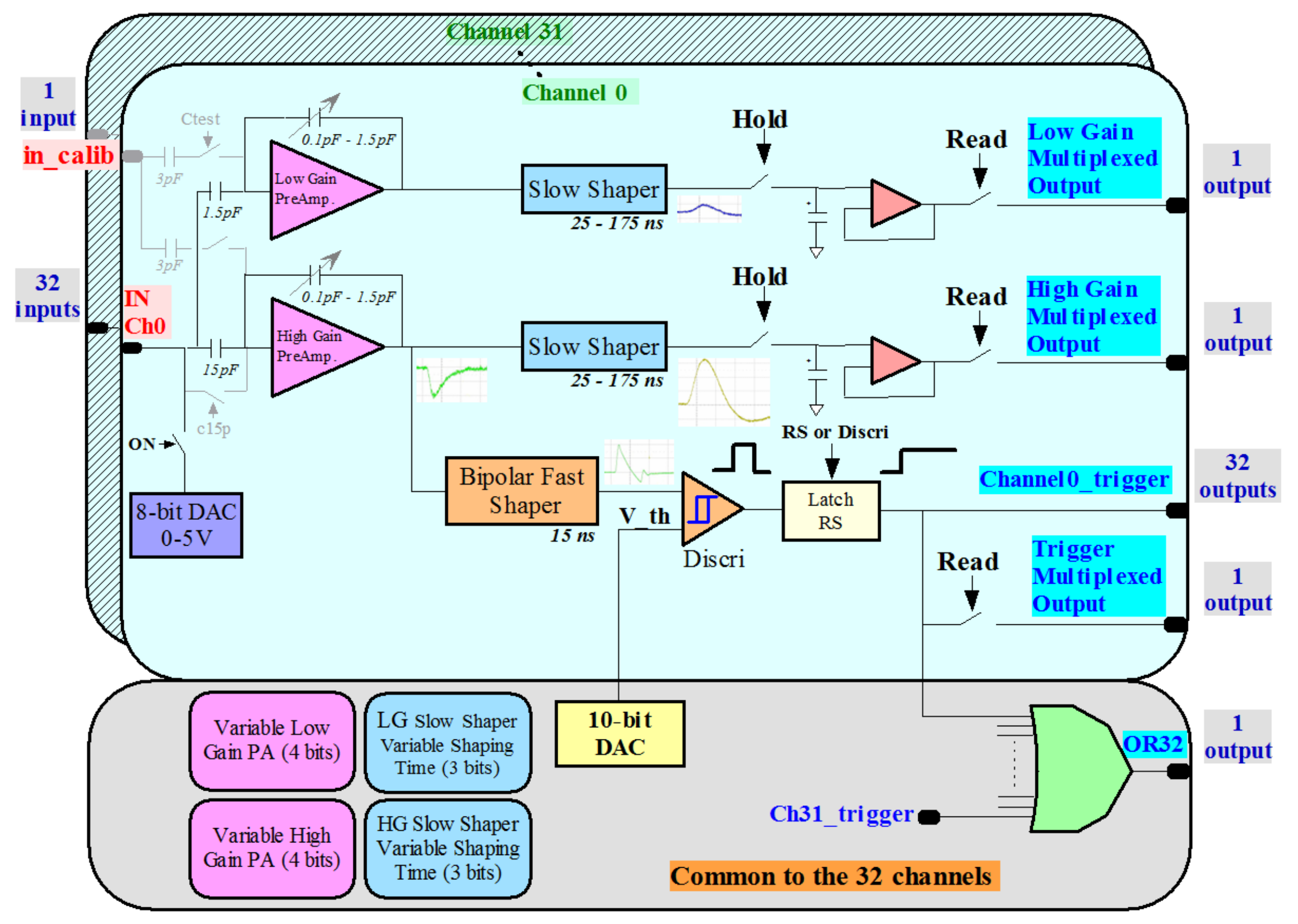}
  \caption{Architecture of the EASIROC chip (baseline).}
  \label{figure.3}
 \end{figure}

The very short duration of the Cherenkov light flashes associated to a gamma-ray, requires an ad-hoc FEE able to provide auto-trigger capability and fast camera pixels read-out.
The main element of the ASTRI SST-2M front-end electronics is the Extended Analogue SiPM
\-In\-te\-gra\-ted Read Out Chip EASIROC \cite{bib:Callier} produced by
Omega\footnote{http://omega.in2p3.fr}.
EASIROC is an ASIC specifically designed to directly interface SiPM detectors.
It works as signal shaper in which the amplitude of the peak contains the value of the input
signal integrated in a specific time window, conversely to the other Cherenkov telescopes that
are based on ASIC that sample the input signal at typical rates of $1~GHz$.
In this way the data throughput is dramatically reduced as the event data are condensed
in just one value.

As sketched in Figure \ref{figure.3}, EASIROC is equipped with 32 input channels, each feeding a double chain of voltage sensitive low noise pre-amplifiers: one chain is intended for Low Gain (LG) and the other one for High Gain (HG).
In this configuration it is possible to measure charges from $160fC$ up to $320pC$; assuming a SiPM gain of $10^{6}$, this corresponds to a range of $1-2000$ photoelectrons, in agreement with the requirements for the maximum number of photoelectrons detected in one logical pixel. The gain of the
pre-amplifiers can be  adjusted ($1-15$ for LG and $10-150$ for HG). Each preamplifier is followed
by a shaper with a selectable shaping time in the range $25-175ns$. At the end, a track-and-hold
circuit is used to catch the peak of the shaper. The signal shaping time is set to $50ns$ to take
into account the signal duration, the introduced noise and the ASIC performance.
The amplitude of all signals sampled by the track-and-hold circuit can be read out by a multiplexer
pair (one for LG and one for HG) and converted by an external ADC. A separate chain, derived from the HG
preamplifier, is implemented to generate a trigger.  The trigger
chain is composed by a fast shaper ($15ns$ shaping time) followed by a discriminator whose threshold
is set by a common 10-bit Digital-to-Analog Converter (DAC).

One of the most important characteristics of EASIROC is that it provides 32 trigger output lines, one per
pixel, allowing the generation of a trigger based on the pattern of the fired pixels
in a very smart and efficient way inside the FPGA of the PDM.
The trigger threshold is chosen in order to have a maximum rate of $\sim300~Hz$ from the whole camera at the focal plane; taking into account that the read-out of all the channels lasts $16\mu$s, this
ensures a dead time of $\sim3\%$.
Moreover, each of the EASIROC input channels is internally connected to a 8-bit DAC, that finely adjusts the SiPM bias voltage in a range $0-4.5V$ allowing gain stabilization and equalization at the level of few percent.
The electronic crosstalk is $<0.3\%$ and the power consumption is lower than $5mW/channel$.

All EASIROC main parameters can be programmed downloading a configuration table through
a slow control serial line. Two EASIROC devices are necessary to read the 64 channels
of each single PDM.

 \begin{figure}[t]
  \centering
  \includegraphics[width=0.455\textwidth]{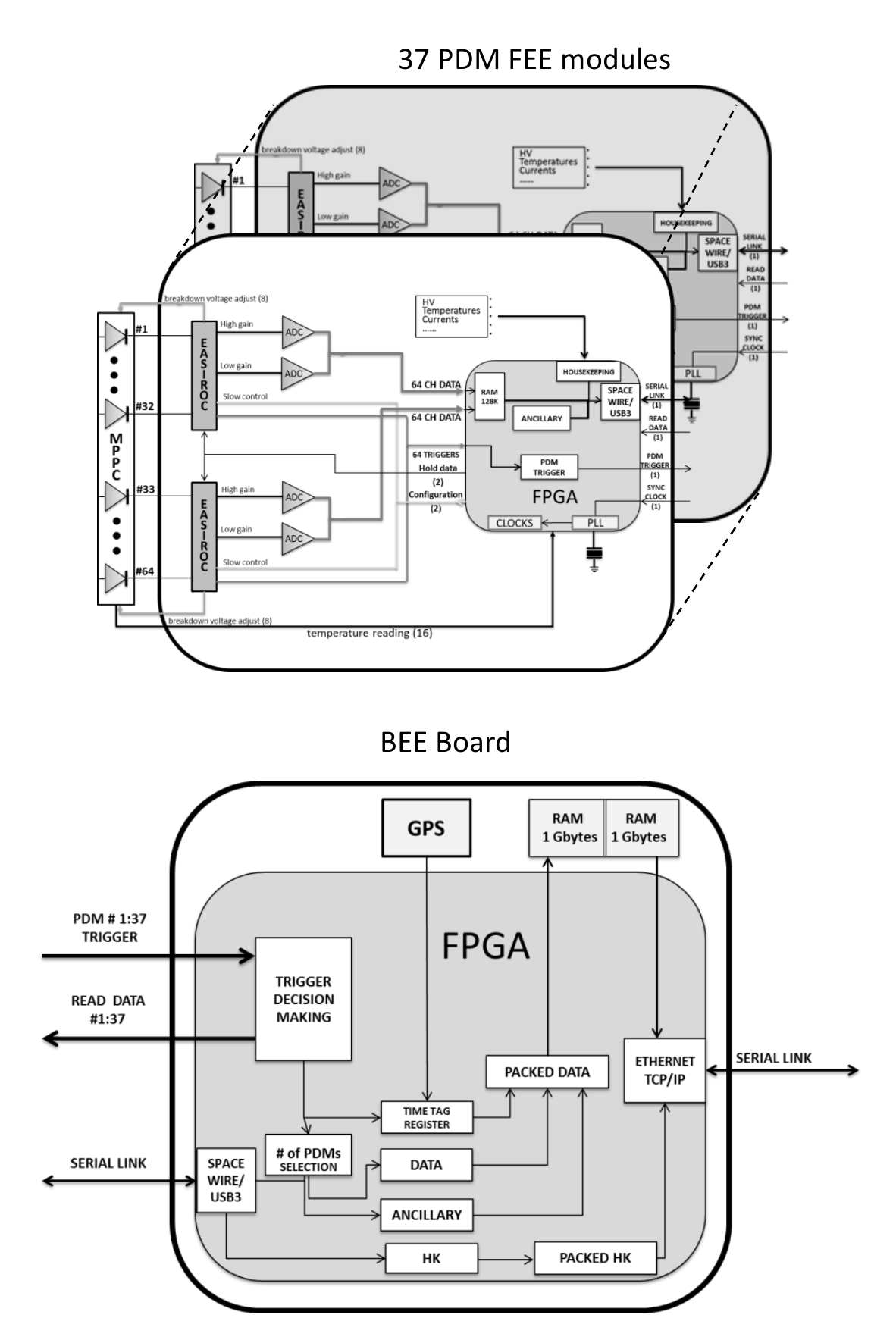}
  \caption{Block diagram of the camera electronics. Top panel: the 37 modules containing the SiPM detectors plus FEE. Bottom panel: structure of the BEE whose tasks are completely managed by his powerful FPGA.}
  \label{figure.4}
\end{figure}

The BEE is based mainly on a powerful FPGA that manages the communication to and from the 37
PDMs and the communication to and from external world. The main tasks of the BEE are: trigger
signals routing, event time tag, storing of the events in a local memory, data packing and
transmission to the external computer, command parser and operating modes management.
Figure~\ref{figure.4} shows the block diagrams of the 37 PDMs and of the BEE.

\section{Lab Measurements}
To verify that the solutions adopted for the camera elec\-tro\-nics and the choice of the
detectors are compliant with the ASTRI SST-2M requirements, plenty of tests were con\-duc\-ted
at the INAF laboratories in Palermo and Catania. The measurements involved mainly the EASIROC
chip and the SiPM detectors \cite{bib:Impiombato, bib:Sottile} and the preliminary results confirm that
they are suitable for such a kind of Cherenkov telescope.
A detailed publication about SiPM and EASIROC characterization will follow in the near future. Among many others, here we show just few features representative of all the tests that were carried out by our group.

\begin{figure}[t]
  \centering
  \includegraphics[width=0.44\textwidth]{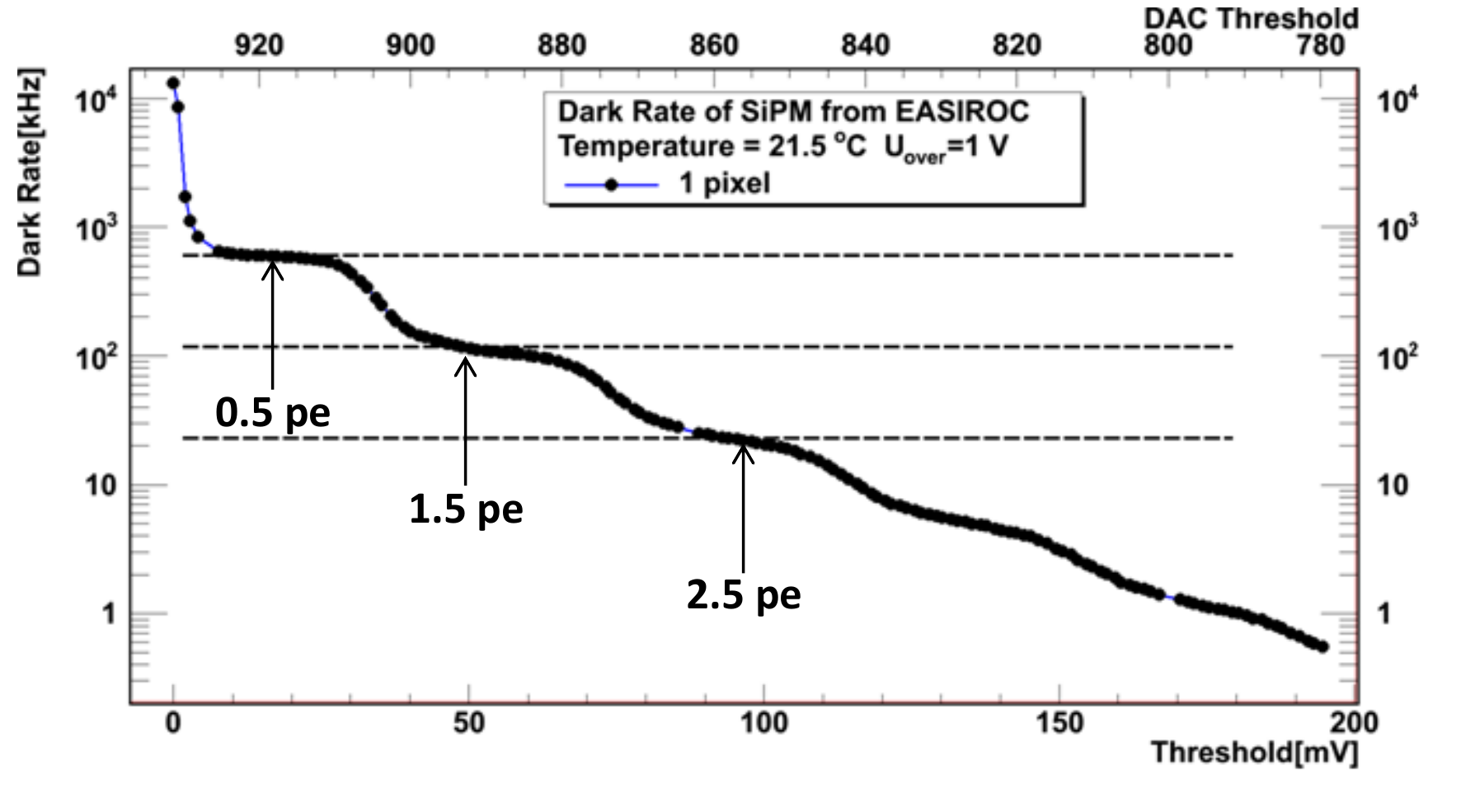}
  \caption{SiPM dark count rate vs threshold. The center of the  plateau in the staircase curve represents the 0.5, 1.5, 2.5 photo-electron (pe) respectively (INAF/IASF Palermo).}
  \label{figure.5}
 \end{figure}

\begin{figure}[t]
  \centering
  \includegraphics[width=0.44\textwidth]{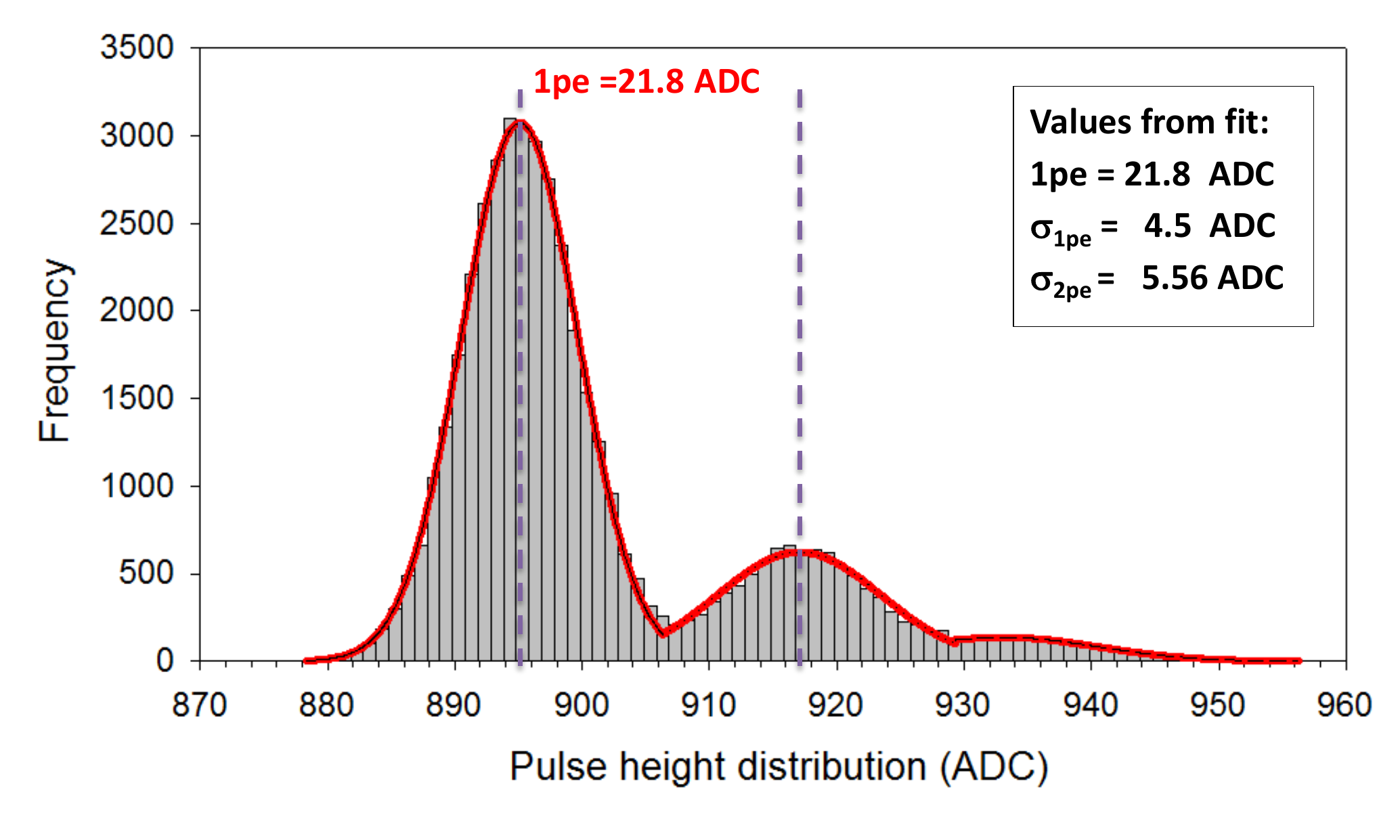}
  \caption{Pulse height distribution of one SiPM pixel.  Separation between peaks represents the pe-equivalent in ADC units (INAF/IASF Palermo).}
  \label{figure.6}
 \end{figure}

\begin{figure}[t]
  \centering
  \includegraphics[width=0.46\textwidth]{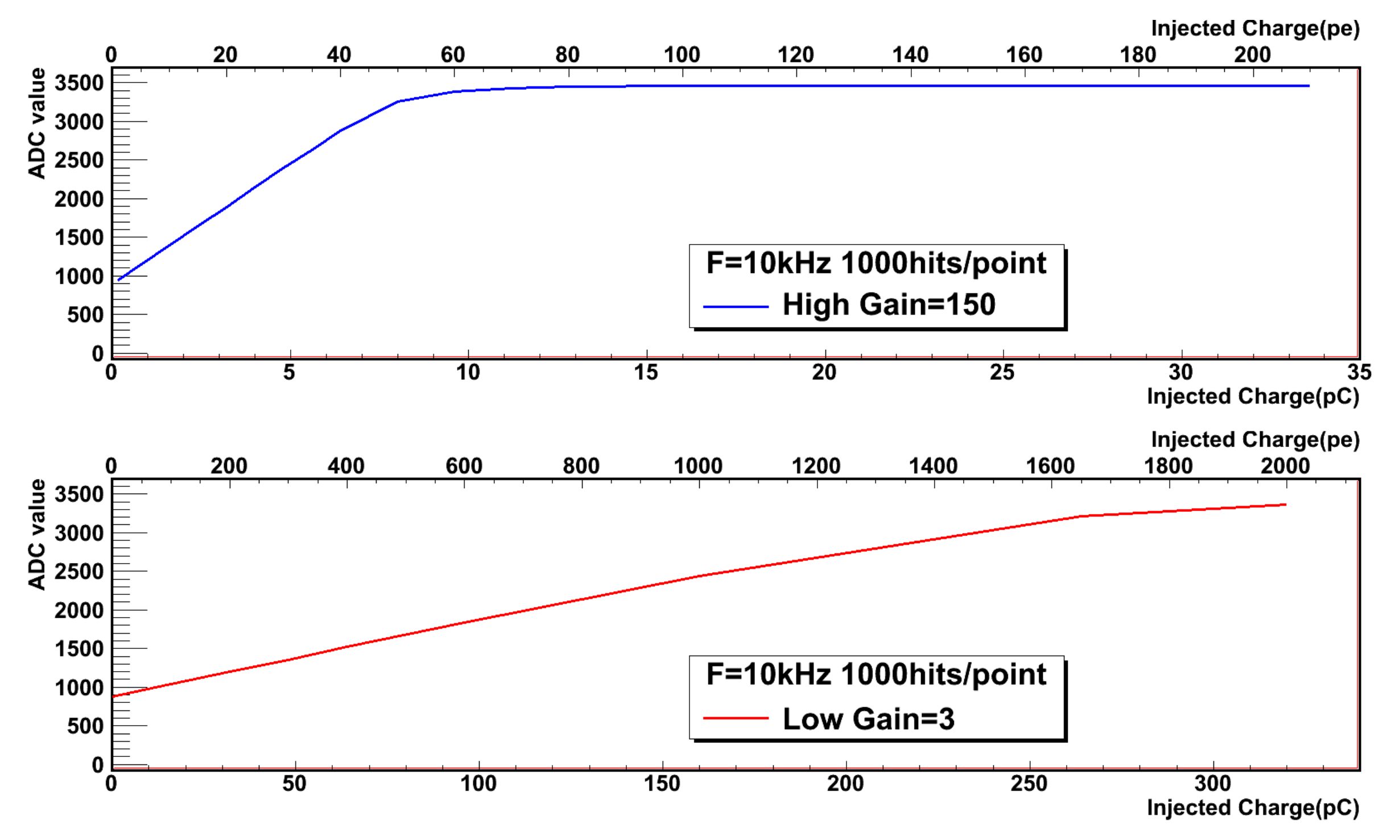}
  \caption{Linearity of EASIROC vs injected charge for HG (top panel) and LG (bottom panel). Note: $320pC$ are equi\-valent to 2000 pe in a SiPM with a gain of $10^6$. (INAF/IASF Palermo).}
  \label{figure.7}
 \end{figure}

\begin{figure}[t]
  \centering
  \includegraphics[width=0.46\textwidth]{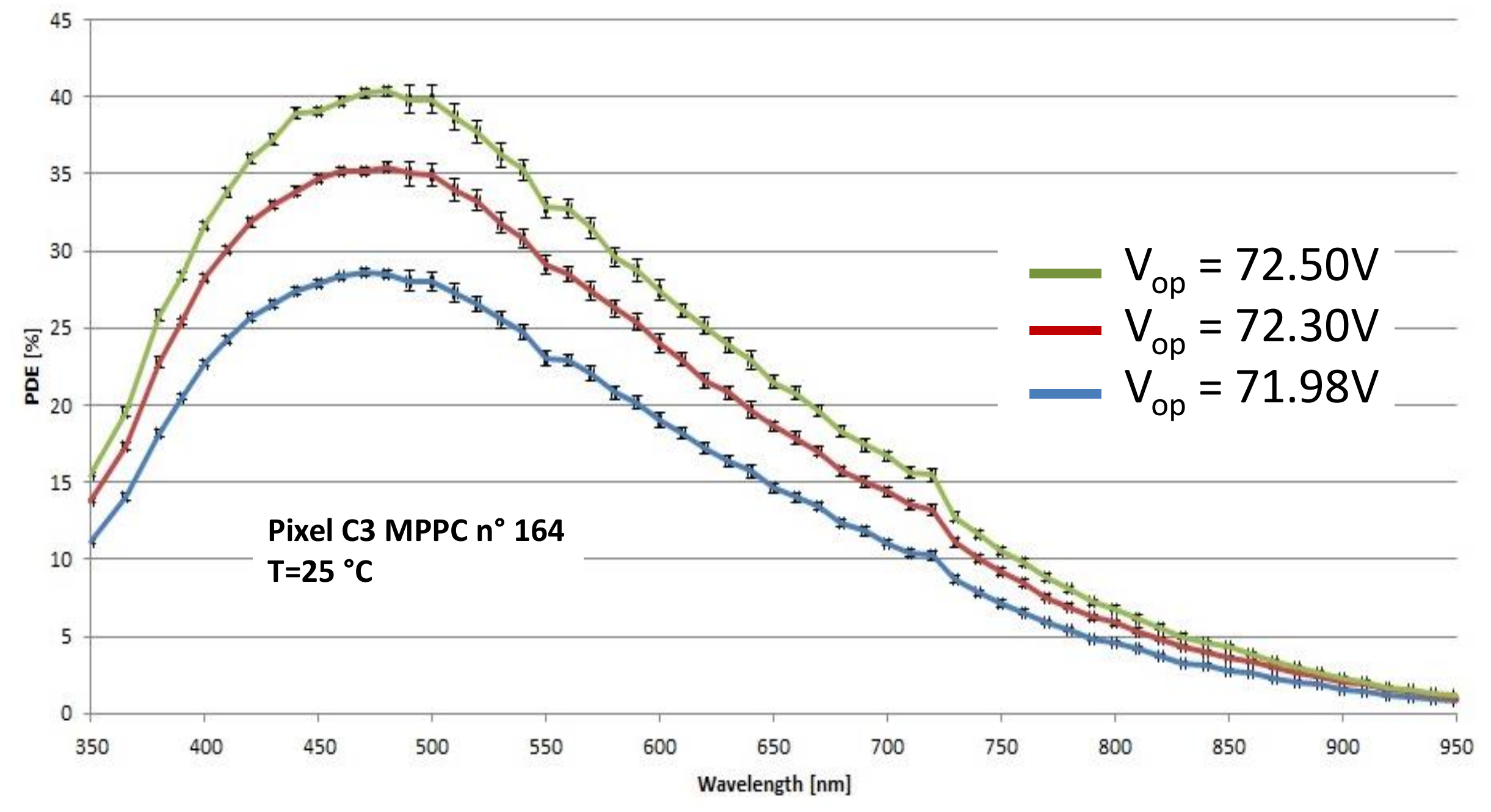}
  \caption{The SiPM detector PDE vs wavelength for three values of operating voltage, $V_{op}$.  (INAF/OA Catania).}
  \label{figure.8}
 \end{figure}

\begin{figure}[t!]
  \centering
  \includegraphics[width=0.45\textwidth]{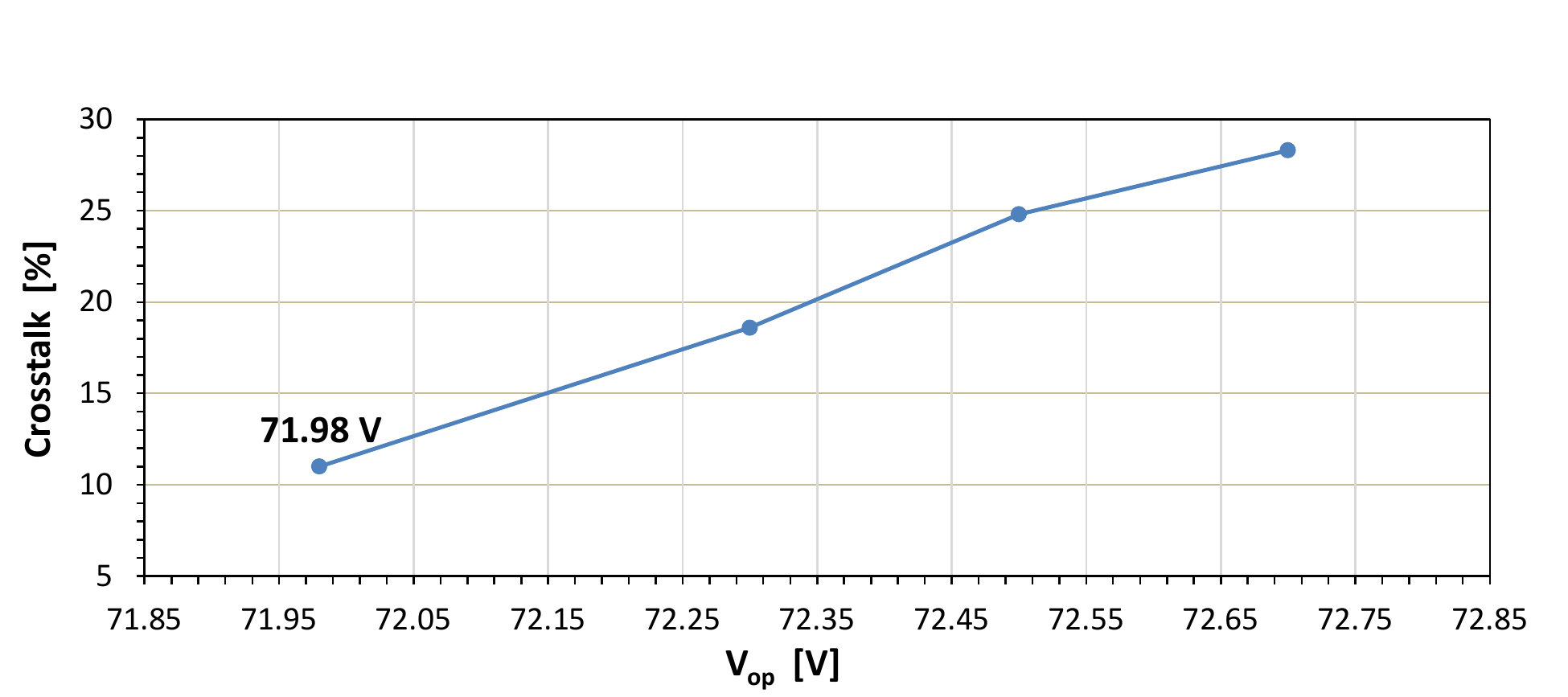}
  \caption{Optical crosstalk effect in SiPM vs operating voltage (INAF/OA Catania).}
  \label{figure.9}
 \end{figure}

Figure~\ref{figure.5} shows the ability of the combined SiPM sensor and  EASIROC in determining the so-called 'staircase curve'. This measurement exploits the adequacy of EASIROC to work in Single Photon Counting mode at low light intensity (SiPM dark current at ambient temperature) and the fine SiPM photon-counting capability. The staircase gives the total dark current rate (about $600kHz$) and  estimates the amount of optical crosstalk of the SiPM (about $16\%$). The probability of optical crosstalk is defined by the rate of dark count events with crosstalk (threshold 1.5 pe ) divided by the total dark count rate (threshold 0.5 fired pe).
The combined performance of SiPM and  EASIROC is also evident in Figure~\ref{figure.6} where is shown the distribution, in ADC units, of the SiPM pulse height (operating in dark current); the two peaks related to the 1st and 2nd pe are well separated with a distance of about 22 ADC determining the pe-equivalent unit.
Figure~\ref{figure.7} shows the response, in ADC counts, of one of the EASIROC channel, as function of the increasing pulse charge for the High Gain and Low Gain chains, respectively; a good linearity up to about 50 pe for the High Gain and 1000 pe for the Low Gain is obtained. Figure~\ref{figure.8} underlines the very good Photon Detection Efficiency (PDE) of the SiPM till the low wavelengths proper of the Cherenkov flashes; the best value of the operating voltage ($V_{op}$) is however a trade-off between PDE and dark counts. The linear behavior of the optical crosstalk vs the operating voltage is shown in Figure~\ref{figure.9} where the value $V_{op}$=71.98V indicates the minimum required to reach the nominal gain declared by Hamamatsu for this SiPM.

\section{Conclusions}
The ASTRI SST-2M telescope is an end-to-end prototype of the Small Size Telescopes for CTA.
The solutions adop\-ted for the camera and for its electronics confirm the compliance with the goal of the project as concerns SiPM si\-gnal processing, pixel dynamic range (1-1000 pe), photon detection efficiency, crosstalk.

At the same time, the outcomes of our lab measurements indicate possible ways of upgrading mainly in the design of the SiPM sensors and in the front-end ASIC which will eventually improve the global performance of the system. Such upgradings are currently in progress in view of the ASTRI mini-array \cite{bib:ASTRI}.

\vspace*{0.5cm}
\footnotesize{{\bf Acknowledgment:}{This work was partially supported by the ASTRI "Flagship Project" financed by the Italian Ministry of Education, University, and Research (MIUR) and led by the Italian National Institute of Astrophysics (INAF). We also acknowledge partial support by the MIUR 'Bando PRIN 2009'.}}

\end{document}